\documentclass[11pt,a4paper]{article}
\usepackage{jcappub}
\usepackage{bm}
\usepackage{latexsym}
\usepackage[latin1]{inputenc}
\usepackage{dcolumn}
\usepackage{amsfonts}
\usepackage{textcomp}
\title{Is Sextans dwarf galaxy in a scalar field dark matter halo?}

\author{V. Lora$^{a}$ and Juan Maga\~na$^{b}$ 
}

\affiliation[a]{Astronomisches Rechen-Institut, Zentrum f\"{u}r Astronomie der Universit\"{a}t Heidelberg, \\
             M\"{o}nchhofstr. 12-14, 69120 Heidelberg, Germany}

\affiliation [b]{Instituto de F\'{\i}sica y Astronom\'{\i}a, Facultad de Ciencias, Universidad de Valpara\'{\i}so,\\ 
Av. Gran Breta\~na 1111, Valpara\'{\i}so, Chile}

\emailAdd{vlora@ari.uni-heidelberg.de}
\emailAdd{juan.magana@uv.cl}

\abstract{
The Bose-Einstein condensate/scalar field dark matter model, considers that the dark matter is 
composed by spinless-ultra-light particles which can be described by a scalar field. This model 
is an alternative model to the $\Lambda$-cold dark matter paradigm, and therefore should be 
studied at galactic and cosmological scales. Dwarf spheroidal galaxies have been very useful when 
studying any dark matter theory, because the dark matter dominates their dynamics. 
In this paper we study the Sextans dwarf spheroidal galaxy, embedded in a scalar field dark matter 
halo. We explore how the dissolution time-scale of the stellar substructures in Sextans, constrain 
the mass, and the self-interacting parameter of the scalar field dark matter boson. We find that for 
masses in the range $(0.12< m_{\phi}<8) \times10^{-22}$~eV, scalar field dark halos without 
self-interaction would have cores large enough to explain the longevity of the stellar substructures 
in Sextans, and small enough mass to be compatible with dynamical limits. If the 
self-interacting parameter is distinct to zero, then the mass of the boson 
could be as high as $m_{\phi}\approx2\times10^{-21}$~eV, but it would correspond to an unrealistic 
low mass fot the Sextans dark matter halo .
Therefore, the Sextans dwarf galaxy could be embedded in a scalar field/BEC dark matter halo with a 
preferred self-interacting parameter equal to zero.}


\begin{document}
\maketitle
\flushbottom
\section{Introduction}
The nature of dark matter (DM) is a puzzle in modern Cosmology. The standard
interpretation supposes that DM is made up of weakly interacting massive 
particles which are non-relativistic at the epoch of decoupling (i.e., cold DM; hereafter CDM). 

Despite the predictions of the CDM are in agreement with several
cosmological observations, it has fundamental inconsistencies which remain to be
solved. A popular example is the well-known  
overpopulation of dark substructure \citep{klypin99}.
Nevertheless, there are many other less-known questions
that the CDM model cannot explain. The expected number of galaxies in the local void 
\citep{peebles10}, is one example.
The observational data at galactic scales seem to disagree with CDM predictions, when comparing 
the density profiles of dark halos predicted in simulations with those derived from observations 
of dwarf spheroidal (dSph) galaxies and Low Surface Brightness galaxies (LSB's). $N$-body 
simulations predict a universal cuspy density profile, while observations indicate that a 
cored halo is preferred in an important fraction of low-mass galaxies \citep{bosch00,kleyna03,blok02}.
This discrepancy is known as the cusp/core problem.

The failures of the CDM model have motivated the study of several DM alternatives.
Lately, a hypothesis that has gained interest is to consider that DM is made up 
bosons described by a real (or complex) scalar field $\Phi$. Such paradigm is 
called the scalar field DM (SFDM) model 
\citep{sin94, ji94, jaeweon96, peebles99, matos00, guzman00, review1, review2}.

In the SFDM model, the scalar field $\Phi$ is minimally coupled to gravity and interacts only 
gravitationally with the baryonic matter. In the early Universe, the scalar 
field is in a thermal bath of temperature $T$. When a critical temperature $T_c$ is 
reached, the scalar field has a symmetry breaking and possibly a phase transition. 
This phase transition can be interpreted as the condensation of the scalar field 
\citep{matos11b, matos_elias12, elias12, elias13}(BEC/SFDM model).
After this stage, the scalar field is driven towards a minimum of the potential. 
Once the scalar field reaches the minimum ($T \ll T_c$), and if the mass of the boson
associated to the scalar field $\Phi$ is greater than the expansion rate of the Universe ($m_{\phi} \gg H$),
then the scalar field has a fast oscillation phase \citep{turner83}. 
At this regime, if the boson mass is in the range $m_{\phi}\sim 10^{-23}-10^{-21}$ eV, 
the SFDM behaves as CDM and their linear perturbations evolve as those in the standard 
CDM model \citep{matos01, matos09, matos11a, magana}. 
Moreover, due to the fact that the effective Jeans length for a scalar field 
depends on the boson mass as $\lambda_{J}\sim m_{\phi}^{-1}$, the mass power 
spectrum has a natural cut-off and the overpopulation of substructures is avoided
in a natural way. 
 
The dynamics of the BEC/SFDM model at cosmological and galactic scales,
have been studied both, theoretically and numerically 
\citep{harkomnras11,harko_core,harko_cosmo,chavanisI,chavanisII,chavanisIII}
to put constraints on the free parameters of the model: mainly the SFDM boson mass.
For example, \cite{rodriguez10}
found that the SFDM model is consistent with the anisotropies of the cosmic 
microwave background radiation (CMB), if the mass of the boson is $m_{\phi}\sim10^{-22}$~eV.
\cite{rindler12}, study the vortex formation in BEC/SFDM halos, including the angular 
momentum, obtaining a window for the mass of the boson of $10^{-21}-10^{-23}$~eV,
for dwarf-galaxy-sized halos.

Recently, \citet{li_rindler} put constraints on a complex SFDM model,
using the effective number of neutrino species during the Big Bang nucleosynthesis,
obtaining $m_{\phi} \geq 2.4 \times10^{-21}\mathrm{eV}/c^{2}$ and 
$9.5 \times 10^{-19}\mbox{eV}^{-1}\mbox{cm}^{3} \leq \lambda/(m_{\phi}c^{2})^{2} \leq 4 \times 10^{-17}\mbox{eV}^{-1}\mbox{cm}^{3}$.

\cite{lora12} use the internal stellar structures of dwarf spheroidal 
(dSph) galaxies to establish a preferred range for the mass $m_{\phi}$ of the bosonic particle. 
They performed $N$-body simulations of the Ursa Minor (UMi) dSph and explored how the dissolution 
time-scale of the cold stellar clump in UMi depends on $m_{\phi}$. They found that for a mass 
in the range of $(0.3<m_{\phi}<1)\times10^{-22}$~eV, the BEC/SFDM model would have large
enough cores to explain the longevity of the cold stellar clump in UMi, and the wide 
distribution of globular clusters in the Fornax dSph. 

On the other hand, \cite{victor_flat} fit the high-resolution rotation curves of a 
sample of $13$ low-surface-brightness galaxies, obtaining a better fit with the 
SFDM/BEC model over the NFW \citep{nfw} profile.
The BEC/SFDM has proved to be a promising DM alternative.
Nevertheless, further tests are needed in galactic systems dominated
by DM, such as the dSph galaxies of the Local Group (LG).

There has been recent evidence of stellar substructure in other dSph galaxy: Sextans. 
\cite{kleyna04} reported the existence of a dissolving cluster at the centre of Sextans. 
Later on, \cite{walker06} detected a region near Sextans core radius that appeared kinematically 
colder than the overall stellar population of Sextans.
Recently, \cite{battaglia11} reported a nine-star group of very metal-poor stars which they
suggest could be in fact the same substructure previously found by \cite{kleyna04}. 

In this work we will perform $N$-body simulations of the Sextans dSph galaxy, embedded in 
a rigid SFDM halo. The survival of the stellar clump in Sextans will give us dynamic constraints 
on the mass $m_{\phi}$ and on the self-interacting parameter $\lambda$. 

The article is organised as follows. In  \S \ref{sec:SFDM} we describe the SFDM model and  briefly review 
the Schr\"{o}dinger-Poisson system. We present the characteristics of 
Sextans and its stellar clump in \S \ref{sec:Sextans}. In \S\ref{sec:Nbody}, we discuss the DM for 
Sextans, the set-up for the $N$-body models, and describe the code used. In \S\ref{sec:results}, we 
describe our results.
Finally, in section \S\ref{sec:conclusions} we discuss the results and give our conclusions.

\subsection{The BEC/SFDM halos}
\label{sec:SFDM}
The BEC/SFDM forms relativistic and Newtonian configurations in equilibrium, which  can be interpreted 
as DM halos. In the relativistic regime, these gravitational structures, described by 
the Einstein-Klein-Gordon (EKG) equations, are known as boson stars (for  complex scalar fields) and 
oscillatons (for real scalar fields). Several numerical studies \citep{ruffini69,seidel91} have shown 
that both structures have a critical mass 
$M_{crit}\sim 0.6 \left( \frac{ m_{P}}{m_{\phi}} \right)\sim10^{12}$~M$_{\odot}$, a typical-galaxy 
size, for an ultra-light boson with mass $m_{\phi}$, and $m_{P}=\sqrt{\hbar c/G}$ the Planck mass.

In this work, we study the dynamics of the Sextans dSph, which is well described by a Newtonian 
gravitational configuration. We model the BEC/SFDM halos, by performing the 
weak field and low velocity limit of the EKG equations, for a complex scalar field $\Phi$, endowed with 
a self-interacting potential $V(\Phi)=m^{2}\Phi^{2}/2+\lambda \Phi^{4}/4$. This leads to the 
Schr\"odinger-Poisson system:

\begin{equation}
 i\hbar\frac{\partial\psi}{\partial t} = -\frac{\hbar^2}{2m_{\phi}} \nabla^2 \psi + 
U m_{\phi} \psi    
+ \frac{\lambda}{2m_{\phi}} |\psi|^2\psi \mbox{ , } 
\label{schroedingerA}
\end{equation}

\begin{equation}
 \nabla^2 U = 4\pi G m_{\phi}^{2} \psi \psi^\ast  \mbox{ . } 
\label{poissonA}
\end{equation}

In the latter equations, $m_{\phi}$ is the mass of the boson associated to $\Phi$. $U$ is the 
gravitational potential produced by the DM density source ($\rho=m_{\phi}^{2}|\psi|^2$), $\lambda$ 
is the self-interacting coupling constant, and the field $\psi$ is related to the relativistic 
field $\Phi$ through $\Phi=e^{-i m_{\phi}c^2t/\hbar}\psi$ \citep{ThesisKevin, ThesisArgelia,friedberg:87}.

To construct spherical BEC/SFDM halos in equilibrium, we assume 
$\psi(r,t)=e^{-i \gamma t/\hbar}\phi(r)$, reducing the Schr\"odinger-Poisson system to an eigenvalue problem for 
$\phi(r)$. 

Using the dimensionless variables
\begin{equation}
\hat{\phi} = \sqrt{4 \pi G\hbar} \frac{\phi}{c^{2}} \nonumber
\end{equation}
\begin{equation}
\hat{r}= \frac{m_{\phi}cr}{\hbar} \nonumber
\end{equation}
\begin{equation}
\hat{t}\equiv \frac{m_{\phi}c^{2}t}{\hbar} 
\end{equation}
\begin{equation}
\hat{U}\equiv \frac{U}{c^{2}} \nonumber
\end{equation}
\begin{equation}
\Lambda = \frac{m_{P}^{2} c \lambda}{ 8\pi m_{\phi}\hbar^3} \nonumber
\end{equation}
\begin{equation}
\hat{\gamma} = m_{\phi}c^2\gamma \mbox{ ,} \nonumber
\end{equation}

\noindent then, the Schr\"odinger-Poisson system now reads as,
\begin{equation}
 \frac{d^2}{d\hat{r}^{2}}(\hat{r}\hat{\phi})= 2 \hat{r} (\hat{U}-\hat{\gamma}) 
+ 2 \hat{r} \Lambda \hat{\phi}{^3} \mbox{ ,}
\label{S-icA} 
\end{equation}
\begin{equation}
 \frac{d^2}{d\hat{r}^{2}}(\hat{r}\hat{U})= \hat{r}\hat{\phi}^2 \mbox{ .}
\label{P-icA}
\end{equation}

Following \cite{guzman04} we construct SFDM halos by obtaining ground state solutions of the system 
(\ref{S-icA}-\ref{P-icA}). The mass of this BEC/SFDM halo can be estimated as
\begin{equation}
M=\int^{\infty}_{0} {\hat{\phi}}^2 \hat{r}^2 d\hat{r}.
\end{equation}
In addition, we define the radius of this configuration as
$r_{95}$, the radius containing $95\%$ of the mass. Note that both properties, the mass and the radius, 
of the scalar halo depend on the boson mass, and the self-interacting term. Thus, to model the Sextans 
halo with several $m_{\phi}$ and $\Lambda$ values, we use the invariance of the Schr\"odinger-Poisson system under the 
following scaling
\begin{equation}
\hat{\phi} \rightarrow  \epsilon^2\hat{\phi}  \nonumber
\end{equation}
\begin{equation}
\hat{U} \rightarrow \epsilon^2\hat{U} \nonumber
\end{equation}
\begin{equation}
\hat{\gamma} \rightarrow \epsilon^2\hat{\gamma} 
\end{equation}
\begin{equation}
\Lambda \rightarrow \epsilon^2\Lambda \nonumber
\end{equation}
\begin{equation}
\hat{r} \rightarrow \epsilon^{-1}\hat{r}  \nonumber
\end{equation}
\begin{equation}
\hat{M} \rightarrow \epsilon \hat{M} \mbox{ .} \nonumber
\end{equation}

\begin{figure*}
  \centering
  \includegraphics[width=0.93\textwidth]{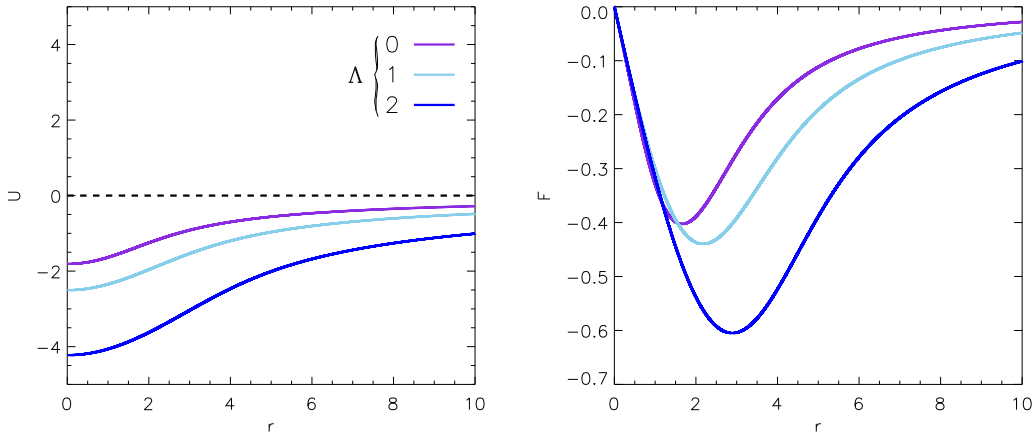}
  \caption{Potential and force associated to the SFDM, for ground states, for the values of the self interacting
parameter $\Lambda=0,1$ and $2$ in dimensionless units.}
\label{fig:U_F}	
\end{figure*}

It is worth mentioning that the excited state solutions of the Schr\"odinger-Poisson system are unstable. Thus, 
we only model the Sextans halo with ground state scalar configurations. 
In Figure \ref{fig:U_F}, we show the potential and the force (in dimensionless units) associated to the SFDM, 
for ground states for three different values of the self interacting parameter $\Lambda$.

Additionally, the Schr\"odinger-Poisson equations could 
be interpreted as the mean-field approximation at zero temperature of the Gross-Pitaevskii-Poisson 
(GPP) system, governing the dynamics of a Newtonian BEC. In the BEC interpretation, $U$ is the trapping 
potential of the BEC, and the coupling constant $\lambda$, is related to $s$-wave scattering length $a$ 
of the bosons. The regime where the self-interacting term strongly dominates the GP equation, is called 
the Thomas-Fermi limit (TFL). For a static BEC in the TFL, the GP reduces to a Lane-Emden (LE) equation, 
which has an analytical solution, when the BEC has a polytropic equation of state (EoS) with index $n=1$. 
In this case, the BEC density profile is $\rho_{BEC}(r) \propto \frac{sin(r/r_{max})}{r/r_{max}}$,
where $r_{max}$, the maximum size of the gravitational structures, depend on $m$ and $\lambda$.
Although these configurations are widely used,
\citet{guzman13_stab} claim that they are unstable 
and dissipate in a short time (see also the discussion by \cite{toth:2014}).
Nevertheless, \cite{deSouza} found that the stability of 
BEC halos in the TFL depend on the mass and the scattering length of the particle.
In this work we do not consider the TFL solutions.


\section{Sextans and its stellar substructures}
\label{sec:Sextans}
The Sextans dSph galaxy satellite of the Milky Way is located at a Galactocentric
distance of $R_{GC}=86$~kpc \citep{mateo98} and it has a luminosity of $
L_{V}=(4.37 \pm 1.69)\times10^{5}$~L$_{\odot}$ \citep{lokas09}. It has a core radius 
of $R_{core}\simeq 0.4$~kpc and a tidal radius of $R_{tidal}\simeq4$~kpc \citep{irwin95}. 
It has a stellar mass of $\sim8.9\pm4.1\times10^5$~M$_{\odot}$ \citep{karlsson12}. The 
values of the mass and luminosity of Sextans give a typical stellar mass-to-light ratio 
of $\Upsilon_{\star} \approx 2$.

Based on the velocity dispersion profile, \cite{walker07} obtain a  
$M/L\sim130$~(M/L)$_{\odot}$. On the other hand, \cite{strigari07} estimate a 
$M/L\sim260$~(M/L)$_{\odot}$, and  \cite{lokas09} obtain a $M/L$ value of 
$\sim96$~(M/L)$_{\odot}$. The latter studies based on Sextans' internal dynamics, suggest 
that the Sextans dwarf is a highly DM dominated dSph.

Sextans is also very interesting because it contains (at least) two stellar substructures.
\cite{kleyna04} found that the velocity dispersion at the centre of Sextans was 
close to zero, and that such a low value of the dispersion was in agreement with 
significant radial gradients in the stellar populations (change in the ratio of red 
horizontal branch stars to blue horizontal branch stars). They suggested that this 
is caused by the sinking and gradual dissolution of a star cluster at the centre 
of Sextans. 

On the other hand, \cite{walker06} presented radial velocities of $294$ possible 
Sextans members. Their data did not confirm \cite{kleyna04}'s report of a 
kinematically distinct stellar population at the centre of Sextans with their 
more complete sample. Instead, they detect a region near Sextans core radius ($\sim 0.4$~kpc) 
kinematically colder than the overall Sextans sample with 95\% confidence.

Lately, \cite{battaglia11} reported nine old stars that share very similar spatial location, 
kinematics, and metallicities, being the substructure's average metallicity $[Fe/H]=-2.6$~dex.
This stellar substructure is consistent with being a remnant of an old stellar cluster, with a
luminosity of $2.2\times10^{4}$~L$_{\odot}$.

The present spatial extent of the substructures is very uncertain. The contours of statistical 
significance for regions of cold kinematics \cite{walker06} show that their stellar substructure is
centred on a location $15$~arcmin north of the Sextans centre and has a radial size of $4$~arcmin 
($\sim100$~pc). On the other hand, the nine innermost metal-poor stars that constitute \cite{battaglia11} 
substructure, are found at $R<0$ \textdegree $.22$, i.e., at $\approx330$~pc.

Since we do not know the orbital parameters of the substructures, we explored different orbits
for the clumps around the Sextans centre. We only know lower limits for the semi-major axes of 
the substructures ($\sim400$ pc for \cite{walker06} substructure, and $\sim200$ pc for \cite{battaglia11}
substructure). Since the substructures are not necessarily on circular orbits, we also consider the
case of eccentric orbits.

\section{The modelling of Sextans}
\label{sec:Nbody}
\subsection{Sextans' Dark Matter component}

\cite{battaglia11} computed the DM mass models for Sextans, based on their 
best-fitting of their observed line-of-sight velocity dispersion profile. 
They found that for a NFW \citep{nfw} DM model, the best fitting model corresponds to 
a concentration $c=10$ and a virial mass $M_V=2.6\times10^9$~M$_{\odot}$.
They also fit their data to a cored DM profile. For the best fitting, they 
obtained a cored radius $r_c=3$~kpc, and a mass within the last measured 
point ($\sim2.3$~kpc assuming a distance to Sextans of $86$~kpc; 
\citeauthor{mateo98} \citeyear{mateo98}) of $M(<R_{last})=4\times10^8$~M$_{\odot}$.

\cite{strigari07,strigari08} compute the Sextans' DM mass within $0.6$~kpc,
$M(<0.6)=0.9\pm 0.4\times10^7$~M$_{\odot}$, for a CDM model. 
\cite{battaglia11} obtained a mass  $M(<0.6)=2 \pm0.6 \times10^7$~M$_{\odot}$, 
for their best-fitting NFW model. They also find an enclosed mass of 
$M(<0.6)=0.9 \pm0.2 \times10^7$~M$_{\odot}$ for their best-fitting cored DM profile, 
very similar to \cite{strigari07}'s mass estimate. Therefore, to construct the Sextans' SFDM 
halo, we impose that the halo mass within $0.6$ is $M(<0.6)=9\times10^6$~M$_{\odot}$.

\begin{figure*}
  \centering
  \includegraphics[width=1.\textwidth]{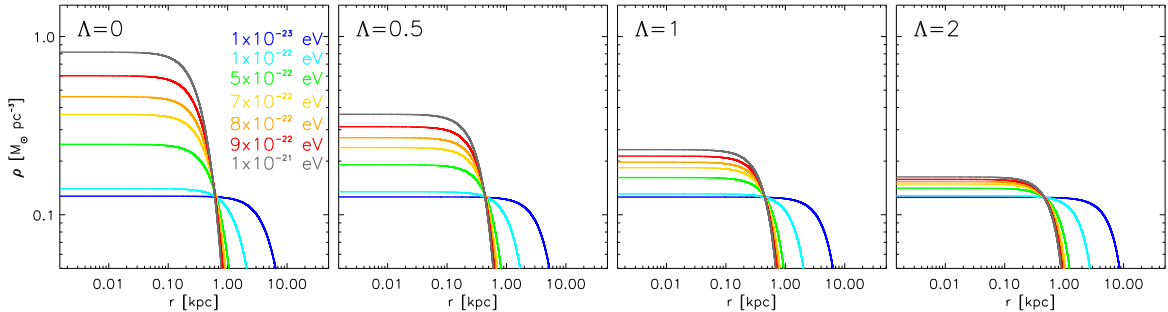}
  \caption{We show the density profile of the SFDM halo for Sextans dwarf galaxy, for
  different values of the mass of the boson $m_{\phi}$, and for the self-interacting parameters
  $\Lambda=0$, $0.5$, $1$ and $2$.}
  \label{fig:FIG1}
\end{figure*}

In Figure~\ref{fig:FIG1}, we show the Sextans' SFDM density profile (for $\Lambda=0$, $0.5$, $1$ 
and $2$) for different mass of the boson $m_{\phi}$. From this Figure, we can see that for
a self interacting parameter $\Lambda=0$, the central density is $\rho_{0}\approx0.8$~M$_{\odot}$~pc$^{-3}$,
for $m_{\phi}=10^{-23}$~eV. For larger values of the self interacting parameter, for example
$\Lambda=2$, the central density for the same mass of the boson $m_{\phi}=10^{-23}$~eV, is
$\rho_{0}\approx0.16$~M$_{\odot}$~pc$^{-3}$. The value of the central density drops when $\Lambda$
increases. 
We define the SFDM core radius, as the radius at which the central density has dropped a factor 
$2$ (see Table~\ref{table:1}). We can see from Figure \ref{fig:FIG1}, that for a fixed mass of
the boson $m_{\phi}$ (say $10^{-21}$~eV), the core radius for $\Lambda=0$ is $r_{core}=0.34$~kpc.
Whereas, for the same $m_{\phi}$ and $\Lambda=2$, the core radius is $r_{core}=0.71$. The value of
the core radius increases when $\Lambda$ increases, for a fixed value of $m_{\phi}$. 

%
Recently, \cite{veraciro14} analysed the Aquarius simulations to characterize the shape of the DM
halos with maximum circular velocities between $8$ and $200$ km/s. They found that DM sub-halos,
comparable to those hosting classical dSph galaxies in the LG, are mildly triaxial with $(b/a)\sim0.75$ and
$(c/a)\sim0.6$ at $r\sim1$~kpc. Therefore, as a first approximation, it is a reasonable assumption
to adopt an spherical BEC/SFDM DM halo for the Sextans dSph.
%

%

\subsection{Sextans' stellar substructures}
\label{sec:clump}
For the initial mass density profile of the stellar clump, we used a Plummer 
\citep{plummer11} mass density profile given by the following equation:

\begin{equation}
\rho_{c}(r) = \rho_{0} \left(1+\left(\frac{r}{r_p}\right)^2\right)^{-5/2} \mbox{ .}
\label{rho_plummer}
\end{equation}

The present spatial extent of the substructures is very uncertain. The contours of statistical 
significance for regions of cold kinematics in \cite{walker06} show, that their substructure is
located at $\sim375$~arcmin north of the Sextans centre, and has a radial size of $\sim100$~pc. 

On the other hand, the nine innermost metal-poor stars that conform the stellar clump 
\citep{battaglia11}, are found at $R<0$\textdegree$.22$ ($\sim330$~pc, if we assume a 
distance to Sextans of $86$~kpc; \citeauthor{mateo98} \citeyear{mateo98}). 
In \citeauthor{battaglia11}'s (\citeyear{battaglia11}) data, there are no metal-poor stars within 
$R<0$\textdegree$.1$, which suggests that the stellar substructure extends in projected galactocentric
radius from $0$\textdegree$.1$~to~$0$\textdegree$.22$ (equivalent to $150-330$~pc). Indicating that, 
in projection, its centre is at $\sim~240$~pc from the centre of Sextans, with a radius (at most) of 
$\sim90$~pc.

We assume that the $V$-band mass-to-light ratio $M/L_{V}$ of the clump is the same as it 
is for the underlying stellar component ($\Upsilon_{\star}=2$). A crude estimate of the 
stellar substructures' total luminosity is $L_c=2.2\times10^4$~L$_{\odot}$ \citep{battaglia11}, 
then, a mass of the clump $M_{c}\simeq 4.4\times10^{4} M_{\odot}$ is obtained. 

We run sets of simulations varying the plummer radius of the clump ($5$, $35$ and $80$~pc). We 
drop the stellar clumps in a circular orbit with a galactocentric distance of $0.4$~kpc, mimicking 
the stellar substructure found by \cite{walker06}. Then, we drop the stellar clumps at a galactocentric
distance of $0.2$~kpc in a circular orbit, representing the stellar substructure found by \cite{battaglia11}.
Finally, we explore the possibility that a stellar substructure could be in an eccentric orbit with an
apocenter distance from the centre of Sextans of $0.4$~kpc, and a pericenter distance of $0.1$~kpc
(which corresponds to an eccentricity of $e=0.6$).

\subsection{The N-body code}
\label{sec:code}
The crossing time is defined as 
\begin{equation}
 t_{cross}=2\pi \frac{R_{c}^{3/2}}{\sqrt{G M_{c}}} \mbox{ ,}
\end{equation}
then the relaxation time can be defined as a function of the crossing time, 
\begin{equation}
 t_{relax} \simeq \frac{0.1 N}{ln N}\times t_{cross} \mbox{ .}
\end{equation}
The relaxation times for both clumps radii are $\gtrsim t_{H}$ therefore the two-body relaxation 
processes can be neglected and the system can be represented as collisionless \citep{binney}.
We simulated the evolution of the Sextans' stellar clump embedded in a rigid SFDM halo potential, 
using the $N$-body code \scriptsize {SUPERBOX} \normalsize \citep{fellhauer00,bien}. 
\scriptsize {SUPERBOX} \normalsize is a highly efficient particle-mesh, collisionless-dynamics 
code with high resolution sub-grids. 
In our case, \scriptsize {SUPERBOX} \normalsize uses three nested grids centred in the 
centre of density of the Sextans' stellar clump. We used $128^3$ cubic cells for each of 
the grids. The inner grid is meant to resolve the inner region of Sextans' clump, and the outer 
grid (with radii of $10$~kpc) resolves the stars that are stripped away 
from Sextans' clump. 
The spatial resolution is determined by the number of grid cells 
per dimension ($N_c$) and the grid radius ($r_{\rm grid}$). Then the side length of one 
grid cell is defined as $l=\frac{2 r_{\rm grid}}{N_c-4}$. For $N_{c}=128$, the resolution
is  $0.5$~pc.
\scriptsize {SUPERBOX} \normalsize integrates the equations of motion with a leap-frog 
algorithm, and a constant time step $dt$. We selected a time step of $dt=0.1$~Myr in our 
simulations in order to guarantee that the energy is conserved better than $1\%$.

\section{Results}
\label{sec:results}

\subsection{The $\Lambda=0$ case}
\label{subsec:lambda0}
We ran $N$-body simulations from $t=0$ to $t=10$~Gyr, of the stellar clump in the Sextans 
dwarf (for three different plummer radius; $5$, $35$ and $80$) embedded in a SFDM halo, 
with a self-interacting parameter $\Lambda=0$, varying the mass of the boson $m_{\phi}$.
As we mentioned before, we selected three different types of orbits; A circular orbit 
at a galactocentric distance of $0.4$~kpc, a circular orbit with a galactocentric distance 
of $0.2$~kpc, and an eccentric orbit with $e=0.6$. All the stellar-clump-models, orbit 
in the $(x,y)$-plane.
In order to quantify the destruction time $\Pi$ of the stellar clump in Sextans, we build 
a map of the surface mass density (in units of M$_{\odot}$~pc$^{-2}$) of the stellar clump 
in the $(x,y)$-plane, for every time $t$ in the simulation. When the surface mass density
falls below the value $\sim1.5$~M$_{\odot}$~pc$^{-2}$, which is the typical surface mass 
density of the underlying stellar component in Sextans, we define that the stellar clump 
is destroyed (i.e., at this value, the particles in the clump would not be recognizable 
form the particles of the main stellar component in Sextans).
%
\begin{figure*}
  \centering
  \includegraphics[width=0.93\textwidth]{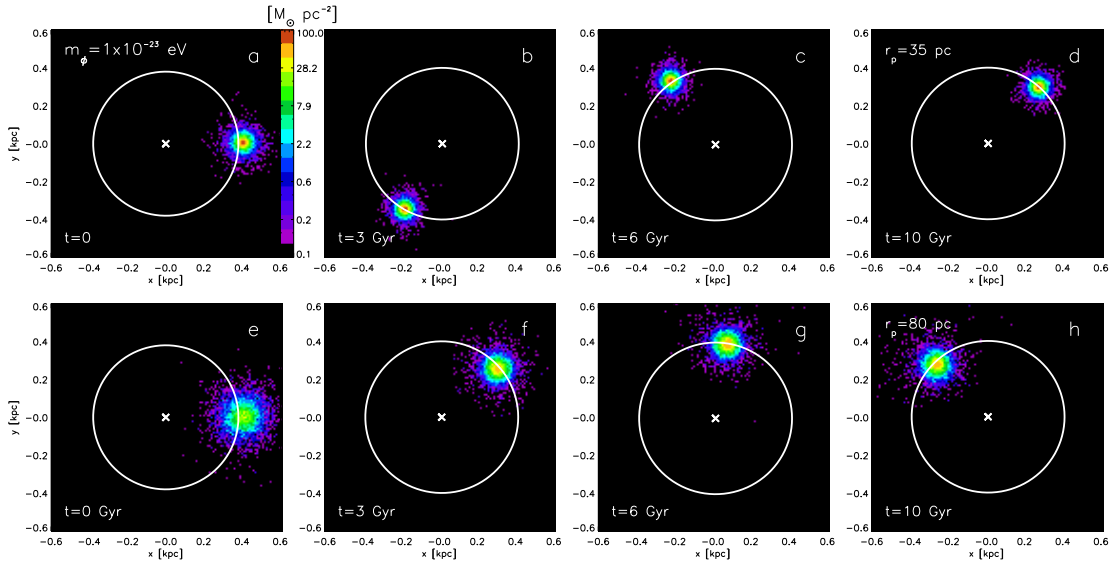}
  \caption{Time evolution ($t=0$, $3$, $6$ and $10$ Gyr) of the stellar clump's surface 
mass density in the Sextans dSph. In the top panels we show the evolution of the clump 
with $r_c=35$ pc, and in the bottom panels we show the time evolution of model with 
$r_c=80$ pc. The white circle shows the initial orbit with $0.4$~kpc radius. The white 
cross marks the centre of Sextans.
We set the clump on a circular orbit in the $(x,y)$-plane at a distance of $r=0.4$~kpc 
from Sextans' centre. The mass of the boson is $m_{\phi}=10^{-23}$~eV and the self interacting 
parameter is $\Lambda=0$, which corresponds to a total mass of the galaxy  
$M=6.28\times10^{9}M_{\odot}$ (see Table \ref{table:1}).}
\label{fig:1e23}	
\end{figure*}
In the top panels of Figure \ref{fig:1e23}, we show the time evolution of the surface 
mass density of the clump at $t=0$, $3$, $6$ and $10$~Gyr, for the model with 
$m_{\phi}=10^{-23}$~eV in a circular orbit at a galactocentric distance of $0.4$~kpc.
The stellar clump has an initial plummer radius, $r_{p}=35$~pc. 
Such a location of the clump,  resembles the stellar substructure found by \cite{walker06}. 
The white line shows the initial orbit of the stellar clump, and the white cross shows the 
centre of Sextans. 

We see that the clump remains intact for $\sim 10$~Gyr. In the
lower panels of Figure \ref{fig:1e23}, we show the time evolution of the stellar clump
with a radius $r_{p}=80$~pc. Also in this extended-radius case, the clump remains unchanged 
for $\sim 10$~Gyr.
The survival of both stellar clumps is a consequence that the SFDM halo has a very large
core ($\sim 5.4$~kpc; see Table \ref{table:1}), that guarantees the survival of the clump
\citep{kleyna03,lora12}. 
In Table~\ref{table:1}, we give the mass of the SF boson, the properties of the SFDM halos, 
the destruction time ($\Pi$) for each clump radius case, and the orbit of the stellar clump.

There is a positive correlation between the size of the SFDM core radius ($r_{core}$), and 
the maximum of the circular velocity ($V_{\rm max}$). Large values of the core are favoured 
to explain the persistence of the stellar clump, but they require very large values of 
$V_{\rm max}$ and thus, the total DM halo mass.
For example, for a SF boson mass of $m_{\phi}=10^{-23}$ eV the total dynamical mass is 
$M\approx6.3\times10^{9}$~M$_{\odot}$, and therefore, $V_{\rm max}\simeq 52$~km~s$^{-1}$.
But a value of $V_{\rm max}\simeq 8-12$~km~s$^{-1}$ is computed for Sextans dwarf 
\citep{zentner03,penarrubia:08,collins:14}, which indicates that the latter DM halo is too 
massive.

Moreover, \cite{battaglia11} suggest that Sextans' DM halo virial mass, derived from the 
assumption of a NFW model, is $\sim2.6\times10^{9}M_{\odot}$. This is a factor $\sim 2$ 
smaller, than the mass of the halo that we find for a $m_{\phi}=10^{-23}$ eV.
\cite{zentner03} point out, that only $5\%$ of the sub-halos in a Milky Way-sized halo have a 
total mass $M>5\times 10^{9}M_{\odot}$. Therefore, $5\times10^{9}$~M$_{\odot}$ is a natural 
first upper limit to the mass of Sextans DM halo \citep{lora12}. 
Adopting the latter maximum value for the mass, we obtain a lower limit to the mass of the 
boson of $m_{\phi} \gtrsim 1.2\times 10^{-23}$~eV.

Both, the shape of the underlying gravitational potential, and the longevity of the clump, depend 
on the mass of the SF boson $m_{\phi}$. Since both, the size of the DM core, and the total mass 
increase when $m_{\phi}$ decreases, the next step is to consider the evolution of the clump in 
models with larger values of $m_{\phi}$. 

We increase the mass of the boson one order of magnitude to $m_{\phi}=10^{-22}$~eV. The stellar clump 
(for the three different radius; $5$, $35$ and $80$~pc) also remains undisturbed for $\sim 10$~Gyr. The 
corresponding core radius and mass are, $1.7$~kpc and $\sim2\times10^8$~M$_{\odot}$, 
respectively (see Table \ref{table:1}). With the latter data, we compute a $V_{\rm max}\simeq 17$~km~s$^{-1}$, 
still large for Sextans, but close enough to set a better lower limit to the mass of the SF boson.

\begin{figure*}
  \centering
  \includegraphics[width=0.93\textwidth]{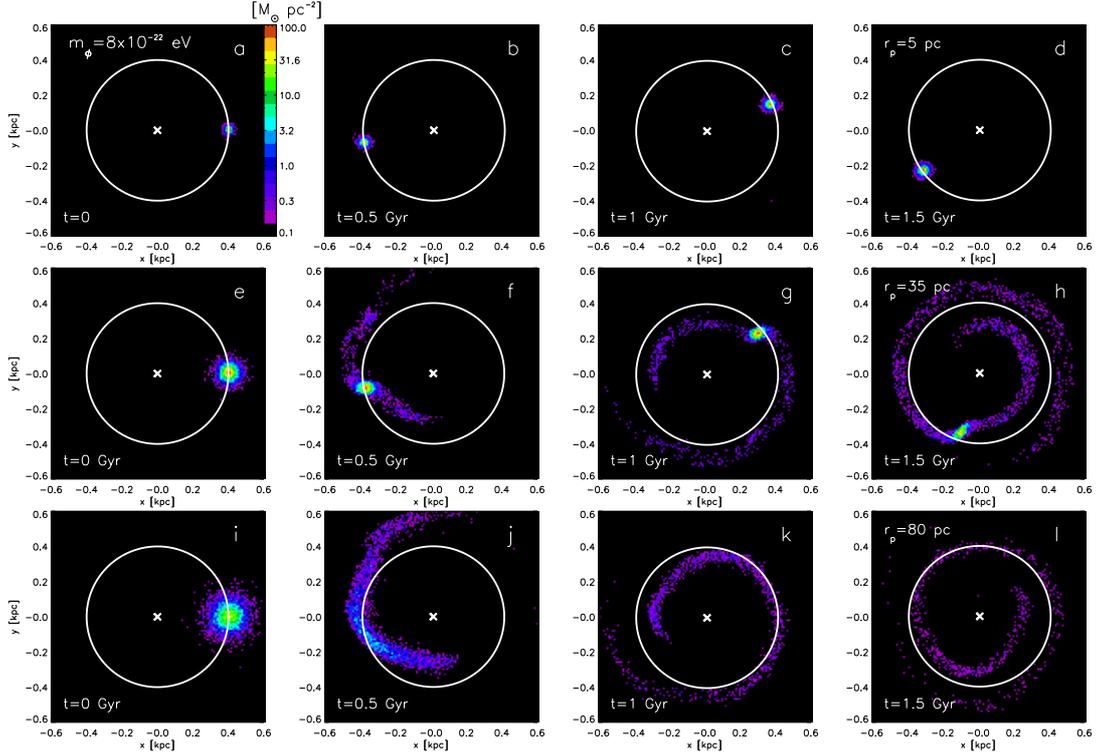}
  \caption{Time evolution ($t=0$, $0.5$, $1$ and $1.5$ Gyr) of the stellar clump's surface 
mass density in the Sextans dSph. In the top panels ($a$-$d$) we show the evolution of the compact 
$r_c=5$~pc clump. In the middle panels ($e$-$h$) we show the time evolution of model with 
$r_c=35$ pc. In the bottom panels ($i$-$l$) we show the time evolution for the extended model
with a radius $r_c=80$ pc.
The white circle shows the clump orbit, and the white cross marks the centre of Sextans.
We set the clump in a circular orbit, in the $(x,y)$-plane, at a galctocentric distance of $0.4$~kpc. 
In this case, the mass of the boson is $m_{\phi}=8\times10^{-22}$~eV and the self interacting parameter is
 $\Lambda=0$, which corresponds to a total mass of the galaxy $M\sim 1.2\times10^{7}M_{\odot}$ (see 
Table~\ref{table:1}).}
\label{fig:8e22}	
\end{figure*}
Next, we rise the value of the mass of the SF boson to $m_{\phi}=5\times10^{-22}$~eV. 
We observe that the clump 
with $r_{p}=35$~pc looses some particles but remains without much 
damage for a Hubble time. On the other hand, for the extended $r_{p}=80$ case, the clump suffers 
a drastic damage, and appears almost destroyed by $\sim10$~Gyr. 
Then, we rise the value of $m_{\phi}$ to $8\times10^{-22}$~eV, the clump (for radius $r_{p}=35$ and $80$~pc) 
is destroyed within $2$~Gyr, due to strong tidal effects (see Figure \ref{fig:8e22}). But a compact stellar 
clump with $r_{p}=5$~pc survives for $10$~Gyr. The $m_{\phi}=8\times10^{-22}$~eV case has a value 
for the core radius of $\sim0.5$~kpc and $V_{\rm max}\simeq 8$~km~s$^{-1}$, which is in agreement with the
value given by \cite{zentner03}.

It has to be noted that the stellar clump with $r_{p}=5$ and $35$~pc remains without being destroyed for
the adopted mass range $m_{\phi}=10^{-23}-10^{-21}$~eV, when the clump is orbiting at a galactocentric
distance of $0.2$~kpc (similar to \citeauthor{battaglia11}'s \citeyear{battaglia11} substructure case). 
The fact that the clump is located well inside the SFDM core radius, guarantees
its longevity. Even the extended $r_{p}=80$~pc case survives for $3$~Gyr, when orbiting so close to the 
centre of the SFDM potential of Sextans.

For the eccentric orbit case, the clumps with $r_{p}=35$ and $80$~pc get destroyed when 
$m_{\phi}=8\times10^{-22}-10^{-21}$~eV. The $r_{p}=5$~pc clump gets destroyed when we raise the
mass of the boson to $m_{\phi}=9\times10^{-22}-10^{-21}$~eV (see Table~\ref{table:1}). 

The survival of the stellar substructures in Sextans set an upper limit to the mass of the boson of 
$m_{\phi} < 8-9\times10^{-22}$~eV. The destruction times, for each of $m_{\phi}$ cases are given in 
Table~\ref{table:1}.
We conclude that the mass of the boson in the $\Lambda=0$ case lays in the range
$10^{-21}< m_{\phi}<8\times10^{-22}$~eV.

\subsection{The small $\Lambda \neq 0$ case}

In the subsection \ref{subsec:lambda0}, we assumed that the boson self-interaction is 
negligible ($\Lambda=0$). In order to see how $m_{\phi}$ depends on self-interaction, we 
explore models with the third term of Equation~\ref{schroedingerA} being distinct from zero 
($\Lambda \neq 0$). We consider only  small values of $\Lambda$ (in dimensionless units $0.5$, 
$1$ and $2$). The parameters of the models are summarized in Table \ref{table:1}. 

For a $\Lambda=0.5$ value, the clump with radius $5$ and $35$~pc remains undestroyed
even when $m_{\phi}=10^{-21}$~eV, for all three different orbital cases: circular orbit with 
a galactocentric distance of $0.2$ and $0.4$~kpc, and eccentric orbit with $e=0.6$. 
Only the extended $r_p=80$~pc case gets destroyed ($\Pi_{80}\approx1.4$ and $0.5$~Gyr) when 
the stellar clump orbits at a galactocentric distance of $0.4$~kpc for $m_{\phi}=8\times10^{-22}$~eV
and  $m_{\phi}=10^{-21}$~eV, respectively. For the other cases (circular orbit at a galctocentric 
distance of $0.2$~kpc and eccentric orbit), even if the stellar clump is so extended, it remains
undestroyed for $10$~Gyr.
Only when we increase the mass of the boson to $m_{\phi}=2\times10^{-21}$~eV, the stellar clump 
gets destroyed for all $r_{p}$, and all orbital cases. This happens because the SFDM core radius 
is too small to guarantee its survival ($\sim0.1$~kpc).

When $\Lambda=1$, the stellar clump only gets destroyed when the mass of the boson has reached 
the high value of $m_{\phi}=2\times10^{-21}$~eV, which corresponds to a DM core radius of $\sim0.2$~kpc). 
The clumps are destroyed earlier, when orbiting at a galactocentric distance of $0.4$~kpc. 
For the orbit with a galactocentric distance of $0.2$~kpc, the stellar clump only gets destroyed for the 
extended stellar clump with $r_{p}=80$~pc. Lastly, when the clump is orbiting the eccentric orbit, the
$r_{p}=35$ and $80$~pc clumps get destroyed within the first Gyr ($0.61$ and $0.46$~Gyr respectively, 
see Table \ref{table:1}). Because the $r_{p}=5$~pc clump is very compact, it overcomes the tidal effect
of the SFDM halo, and gets destroyed later at $\sim3.77$~Gyr.

\begin{figure*}
  \centering
  \includegraphics[width=0.93\textwidth]{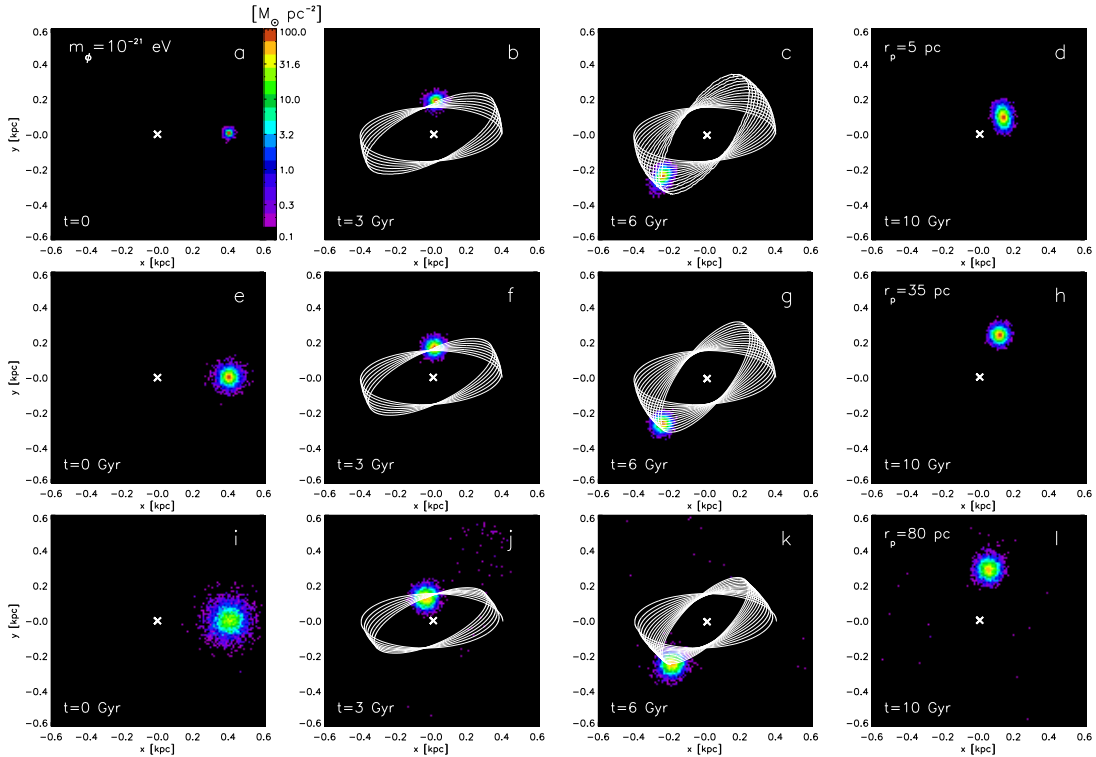}
  \caption{Time evolution ($t=0$, $3$, $6$ and $10$ Gyr) of the stellar clump's surface 
mass density in the Sextans dSph for a $m_{\phi}=10^{-21}$~eV, and with a self interacting parameter of
 $\Lambda=2$. In the top panels ($a$-$d$) we show the evolution of the compact 
$r_c=5$~pc clump. In the middle panels ($e$-$h$) we show the time evolution of model with 
$r_c=35$ pc. In the bottom panels ($i$-$l$) we show the time evolution for the extended model
with a radius $r_c=80$ pc.
The white line shows the clump orbit in the $(x,y)$-plane (with apocenter at $0.4$~kpc and pericenter 
at $0.1$~kpc), and the white cross marks the centre of Sextans.}
\label{fig:21_L_2}	
\end{figure*}
When  $\Lambda=2$, we observe that the clump never gets destroyed, for any radius $r_{p}$ case, 
or any orbital case. In Figure~\ref{fig:21_L_2}, we show the case of the stellar clump embedded in 
a SFDM halo with $m_{\phi}=10^{-21}$~eV. For such a mass of the boson, and self-interacting parameter,
the SFDM core radius is $0.7$~kpc, guaranteeing the survival of the clump. But it has to be noted, that 
in order to have a total mass of the DM halo $\gtrsim 10^{8}$~M$_{\odot}$, as suggested by 
\cite{strigari07}, a $m_{\phi}\approx10^{-22}$~eV is needed for a self interacting 
parameter $\Lambda=2$.

For the same value of $m_{\phi}$, the masses and core radii of the DM halos increase when 
$\Lambda$ increases (see Table \ref{table:1}). For example, for a boson mass of $10^{-21}$~eV, 
the SFDM core radius is $0.35$, $0.47$ and $0.71$ for $\Lambda=0.5$, $1$ and $2$, respectively. 
Then, the permitted window for the mass $m_{\phi}$ of the bosonic particles is shifted to 
larger values ($m_{\phi}\approx10^{-21}$~eV).
We observe from our results, that the clump must be embedded in a SFDM halo with a mass 
$M\approx10^7$~$M_{\odot}$, a ``size'' of the DM halo of $r_{95}\approx0.8$~kpc, and a core radius
of $\approx0.4$~kpc, in order to guarantee the survival of the stellar substructures.
 

\section{Conclusions}
\label{sec:conclusions}
In this work we consider an alternative to the CDM model, where ultra-light bosons 
are the main components of the DM halos. We constrain the mass of the ultra-light 
bosons (the SFDM particle) using as a tool the stellar substructures found in the
Sextans dwarf galaxy. Using $N$-body simulations, we found that the survival of the
stellar substructures is only possible if $m_{\phi}<8\times10^{-22}$~eV 
for a self-interacting parameter $\Lambda=0$.
By imposing a realistic upper limit on the dynamical mass of Sextans,
we place a lower limit of $1.2\times10^{-23}$~eV. 
Therefore, we have a possible mass window for the SF boson of 
$1.2\times10^{-23}$~eV$m_{\phi}$<$8\times10^{-22}$~eV.
These constraints imply
SFDM halos with masses between $10^{7}-10^{9}~M_{\odot}$, maximum circular velocities
between $48-8$~km~s$^{-1}$, and sizes between $\sim1-10$~kpc.

For the SFDM halos, where we include the self-interacting parameter, the upper limit grows
with $\Lambda$. For example, for $\Lambda=2$, the halos made up by bosons with a mass as high 
as $m_{\phi}=2\times10^{-21}$~eV, accounts for the observed internal Sextans dynamics 
($v_c\approx8$~km~s$^{-1}$), but the corresponding DM halo mass 
($1.3\times10^{7}$~M$_{\odot}$) would be to low for Sextans dwarf.


The preferred range for the mass of the boson, found in this work, derived from the
dynamics of Sextans, is compatible with those given by other authors to ameliorate 
the problem of over-abundance of dark substructure \citep{hu00,matos01}. It is also 
in agreement with the mass range for the boson mass based on the dynamics of the UMi
and the Fornax dSph galaxies \citep{lora12}.

The BEC/SFDM model has several challenges to overcome mainly
at galactic scales and related to the baryonic dynamics.
For example, \cite{alma} found discrepancies of one order of magnitude
in the size of TFL BEC/SFDM halos estimated from dwarf galaxy dynamics and those derived from 
other galactic systems (strong lensing, rotation curves). 
Nevertheless, \citet{victor_sl} can reconcile the discrepancy
when the finite temperature corrections in BEC/SFDM halos are considered \citep{victor_tcores}. 
The BEC/SFDM configurations in the TFL for ultra-light bosons are numerically unstable.
However, the problem could be solved if the BEC/SFDM halos have angular momentum \citep{guzman13_L}
or if the boson mass and the scattering length have an appropriate value \cite{deSouza}. 
The BEC/SFDM is a viable model to explain the nature of the DM,
at cosmological and galactic scales, and therefore
should be tested in more astrophysical systems.

\section*{Acknowledgments}

We would like to thank Andreas Just, for his useful comments and suggestions.
V.L. gratefully acknowledges support from the FRONTIER grant, and HB-L.
J.M. acknowledges support from ESO Comit\'e-Mixto 2010, and Gemini Fondecyt 32130024. 
\newpage

%
\begin{table*}
 \centering
  \footnotesize 
 \caption{Destruction times ($\Pi$) of the stellar clump in Sextans for the 
$\Lambda=0$, $0.5$, $1$ and $2$ case, for different values of $m_{\phi}$. The sub-index in $\Pi$ 
corresponds to the plummer radius of the clump ($5$, $35$ and $80$~pc). 
The SF boson mass, $\epsilon$, the total mass $M$; the radius at which $95\%$ of the total mass 
is contained, $r_{95}$; and the core radius, $r_{core}$; are also given for each of the models.}
 \medskip
 \begin{tabular}{@{}cccccccccc@{}}
 \hline
 $\Lambda$ & m$_{\phi}$     & $\epsilon$  &             M          &   r$_{95}$   & r$_{core}$ & $\Pi_{5}$ & $\Pi_{35}$ & $\Pi_{80}$ & Orbit\\
           & ($10^{-22} eV$)& ($10^{-5}$) &  ($10^{8}$~M$_{\odot}$)&      (kpc)   &    (kpc)   &  (Gyr)     &   (Gyr)    &   (Gyr)   &  type \\
  \hline
  &  &  &  \\
0 & 0.1 & 22.7996 & 62.8  & 10.6555& 5.4345& $>$10 &$>$10 &$>$10     & circular, x=200 pc \\
0 & 0.1 & 22.7996 & 62.8  & 10.6555& 5.4345& $>$10 &$>$10 &$>$10     & circular, x=400 pc \\
0 & 1.0 &  7.3881 &  2.035&  3.2882& 1.6770& $>$10 &$>$10 &$>$10     & circular, x=200 pc    \\
0 & 1.0 &  7.3881 &  2.035&  3.2882& 1.6770& $>$10 &$>$10 &$>$10     & circular, x=400 pc    \\
0 & 5.0 &  3.8120 &  0.21 &  1.2746& 0.6501& $>$10 &$>$10 &$>$10     & circular, x=200 pc \\
0 & 5.0 &  3.8120 &  0.21 &  1.2746& 0.6501& $>$10 &$>$10 &$\sim 6$  & circular, x=400 pc \\
0 & 7.0 &  3.5477 &  0.139&  0.9782& 0.4989& $>$10 &$>$10 &$>$10     & circular, x=200 pc\\
0 & 7.0 &  3.5477 &  0.139&  0.9782& 0.4989& $>$10 &$>$10 &$\sim 0.5$& circular, x=400 pc\\

0 & 8.0 &  3.5152 & 0.121 &  0.8638& 0.4406& $>$10 &$>$10 &$>$10     & circular, x=200 pc \\
0 & 8.0 &  3.5152 & 0.121 &  0.8638& 0.4406& $>$10 & $\sim 1.7$ &$\sim 0.35$& circular, x=400 pc \\
0 & 8.0 &  3.5152 & 0.121 &  0.8638& 0.4406& $>$10 & $\sim4.53$ &$\sim 0.87$& excentric, e=0.6  \\

0 & 9.0 &  3.5452 & 0.108 &  0.7614& 0.3883& $>$10 & $>$10&$>$10     & circular, x=200 pc \\
0 & 9.0 &  3.5452 & 0.108 &  0.7614& 0.3883& $\sim7.72$   & $\sim 0.65$&$\sim 0.2$ & circular, x=400 pc \\
0 & 9.0 &  3.5452 & 0.108 &  0.7614& 0.3883& $\sim8.08$   & $\sim 1.32$&$\sim0.74$ & excentric, e=0.6 \\

0 & 10  &  3.6305 &  0.1  &  0.6691& 0.3412& $>$10 &$>$10 &$\sim3$   & circular, x=200 pc \\
0 & 10  &  3.6305 &  0.1  &  0.6691& 0.3412& $\sim5.3$    &$\sim 0.4$ &$\sim0.15$ & circular, x=400 pc \\
0 & 10  &  3.6305 &  0.1  &  0.6691& 0.3412& $\sim6$      &$\sim0.97$ &$\sim 0.58$ & excentric, e=0.6 \\

  &  &  &  &  & \\
  \hline
   &  &  &  &  & \\
0.5 & 0.1 &  22.7423 & 95.79   &  11.2252 & 4.5685& $>$10 & $>$10 & $>$10& circular, x=200 pc\\
0.5 & 0.1 &  22.7423 & 95.79   &  11.2252 & 4.5685& $>$10 & $>$10 & $>$10& circular, x=400 pc\\
0.5 & 1.0 &   7.3124 &  3.08   &   3.4911 & 1.4208& $>$10 & $>$10 & $>$10& circular, x=200 pc\\
0.5 & 1.0 &   7.3124 &  3.08   &   3.4911 & 1.4208& $>$10 & $>$10 & $>$10& circular, x=400 pc\\
0.5 & 5.0 &   3.5684 &  0.3006 &   1.4308 & 0.5823& $>$10 & $>$10 & $>$10& circular, x=200 pc\\
0.5 & 5.0 &   3.5684 &  0.3006 &   1.4308 & 0.5823& $>$10 & $>$10 & $>$10& circular, x=400 pc\\
0.5 & 7.0 &   3.1875 &  0.1918 &   1.1441 & 0.4656& $>$10 & $>$10 & $>$10& circular, x=200 pc\\
0.5 & 7.0 &   3.1875 &  0.1918 &   1.1441 & 0.4656& $>$10 & $>$10 & $>$10& circular, x=400 pc\\

0.5 & 8.0 &   3.0788 &  0.1621 &   1.0364 & 0.4218& $>$10 & $>$10 & $>$10& circular, x=200 pc\\
0.5 & 8.0 &   3.0788 &  0.1621 &   1.0364 & 0.4218& $>$10 & $>$10 & $\sim 1.4$& circular, x=400 pc\\
0.5 & 8.0 &   3.0788 &  0.1621 &   1.0364 & 0.4218& $>$10 & $>$10 & $>$10& excentric, e=0.6\\

0.5 & 10  &   2.9701 &  0.1251 &   0.8595 & 0.3498& $>$10 & $>$10 & $>$10& circular, x=200 pc\\
0.5 & 10  &   2.9701 &  0.1251 &   0.8595 & 0.3498& $>$10 & $>$10 & $\sim 0.5$& circular, x=400 pc\\
0.5 & 10  &   2.9701 &  0.1251 &   0.8595 & 0.3498& $>$10 & $>$10 & $\sim 1.38$& excentric, e=0.6\\

0.5 & 20  &   4.2782 &  0.0901 &   0.2983 & 0.1214& $\sim1.75$ & $\sim0.25$ & $\sim0.13$& circular, x=200 pc\\
0.5 & 20  &   4.2782 &  0.0901 &   0.2983 & 0.1214& $\sim2.2$ & $\sim 0.12$& $\sim0.05$& circular, x=400 pc\\
0.5 & 20  &   4.2782 &  0.0901 &   0.2983 & 0.1214& $\sim1.6$ & $\sim 0.3$& $\sim 0.27$& excentric, e=0.6\\

    &  &  &  &  & \\
  \hline
    &  &  &  &  & \\
1 & 1.0 &  7.2623 &  4.693  & 3.7225 &  1.7387 & $>$10 &$>$10 & $>$10& circular, x=200 pc\\
1 & 1.0 &  7.2623 &  4.693  & 3.7225 &  1.7387 & $>$10 &$>$10 & $>$10& circular, x=400 pc\\

1 & 10  &  2.6492 &  0.1712 & 1.0204 &  0.4766 & $>$10 &$>$10 & $>$10& circular, x=200 pc\\
1 & 10  &  2.6492 &  0.1712 & 1.0204 &  0.4766 & $>$10 &$>$10 & $>$10& circular, x=400 pc\\
1 & 10  &  2.6492 &  0.1712 & 1.0204 &  0.4766 & $>$10 &$>$10 & $>$10& excentric, e=0.6\\

1 & 20  &  2.8164 &  0.091  & 0.4799 &  0.2241 & $>$10 &$>$10 & $\sim1.5$& circular, x=200 pc\\
1 & 20  &  2.8164 &  0.091  & 0.4799 &  0.2241 & $\sim 3.43$ &$\sim 0.1$  &$\sim0.09$& circular, x=400 pc\\
1 & 20  &  2.8164 &  0.091  & 0.4799 &  0.2241 & $\sim 3.77$ &$\sim 0.61$ &$\sim0.46$& excentric, e=0.6\\

    &  &  &  &  & \\
  \hline
    &  &  &  &  & \\
2 & 1.0 &  7.2202 &   0.9723 &  4.2634 & 2.3910 & $>$10 &$>$10 & $>$10& circular, x=200 pc\\
2 & 1.0 &  7.2202 &   0.9723 &  4.2634 & 2.3910 & $>$10 &$>$10 & $>$10& circular, x=400 pc\\

2 & 10  &  2.4260 &   0.3267 &  1.2688 & 0.7116 & $>$10 &$>$10 & $>$10& circular, x=200 pc\\
2 & 10  &  2.4260 &   0.3267 &  1.2688 & 0.7116 & $>$10 &$>$10 & $>$10& circular, x=400 pc\\
2 & 10  &  2.4260 &   0.3267 &  1.2688 & 0.7116 & $>10$ &$>10$ & $>10$& excentric, e=0.6\\

2 & 20  &  1.9124 &   0.1288 &  0.8048 & 0.4513 & $>$10 &$>$10 & $>$10& circular, x=200 pc\\
2 & 20  &  1.9124 &   0.1288 &  0.8048 & 0.4513 & $>$10 &$>$10 & $>$10& circular, x=400 pc\\
2 & 20  &  1.9124 &   0.1288 &  0.8048 & 0.4513 & $>$10 &$>$10 & $>$10& excentric, e=0.6\\

   &  &  &  &  & \\
  \hline
  \end{tabular}
  \label{table:1}
 \end{table*}

\def \prl {Phys.\ Rev.\ Lett. }
\def \ijmpd {Int.\ J.\ Mod.\ Phys.\ D }
\def \PRD {Phys.\ Rev.\ D. }
\def \mnras {Mon.\ Non.\ Roy.\ Astron.\ Soc. }
\def \plb {Phys.\ Lett.\ B } 
\def \ApJ {Astrophys.\ J. }
\def \AA {Astron.\ Astrophys. }
\def \AJ {Astron.\ J. }
\def \JCAP {J. Cosmol. Astropart. Phys. }

%

\end{document}